\documentclass[aps,pra,reprint,superscriptaddress,raggedbottom,floatfix]{revtex4-1}
\usepackage{amsmath}
\usepackage{amssymb}
\usepackage[pdftex]{graphics}
\usepackage{epsfig}
\usepackage{xspace}

\usepackage{MnSymbol}
\usepackage{pifont}
\usepackage[geometry]{ifsym} % For referring to certain symbols used in graphs

\newcommand{\beq}{\begin{equation}}
\newcommand{\eeq}{\end{equation}}

\usepackage{tabularx}
\newcolumntype{C}[1]{>{\centering}p{#1}}
\newcolumntype{L}[1]{>{\raggedright}p{#1}}
\newcolumntype{R}[1]{>{\raggedleft}p{#1}}

\newcommand{\usr}[2]{\ensuremath{#1_{\mathrm{#2}}}} % helper for upright subscripts

\newcommand{\unit}[1]{\ensuremath{\;{\rm #1}}}
\newcommand{\units}[2]{\ensuremath{\;{\rm #1}^{#2}}}
\newcommand{\unitd}[4]{\ensuremath{\:{\rm #1}^{#2}\,{\rm #3}^{#4}}}

\newcommand{\mval}[1]{\left\langle #1 \right\rangle} 
\newcommand{\mhyp}{\mbox{-}} 

\newcommand{\nup}{\ensuremath{\usr{\nu}{p}}}
\newcommand{\omp}{\ensuremath{\usr{\omega}{p}}}
\newcommand{\epsb}{\ensuremath{\usr{\epsilon}{br}}} 
\newcommand{\epsz}{\ensuremath{\usr{\epsilon}{0}}}

\newcommand{\meff}{\ensuremath{\usr{m}{eff}}}
\newcommand{\mel}{\ensuremath{\usr{m}{e}}}
\newcommand{\mDOS}{\ensuremath{\usr{m}{de}}}

\newcommand{\ud}{\mathrm{d}} 
\newcommand{\nuex}{\usr{\nu}{ex}}
\newcommand{\Tfwhm}{\usr{T}{fwhm}} 
\newcommand{\lamx}{\usr{\lambda}{ex}} 

\newcommand{\Nex}{\usr{N}{ex}} 
\newcommand{\Nexthr}{\usr{N}{ex}^{\mathrm{thr}}} 
 
\newcommand{\absex}{\usr{\alpha}{ex}} 
\newcommand{\taueph}{\usr{\tau}{e\mhyp ph}} 
 
\newcommand{\Gameph}{\usr{\Gamma}{e\mhyp ph}} 
\newcommand{\Gameh}{\usr{\Gamma}{e\mhyp h}}

\newcommand{\Mnu}{\ensuremath{M(\nu)}} 
\newcommand{\tmu}{\ensuremath{\bar {t}(\nu)}} 
\newcommand{\tmuz}{\ensuremath{\bar{t}_0(\nu)}}

\newcommand{\taus}{\usr{\tau}{s}} 
\newcommand{\tausz}{\usr{\tau}{s0}} 
\newcommand{\psec}{\unit{ps}} 
\newcommand{\fsec}{\unit{fs}} 
\newcommand{\thz}{\unit{THz}} 
\newcommand{\mm}{\unit{mm}} 
\newcommand{\mum}{\unit{\mu m}} 
\newcommand{\nm}{\unit{nm}} 
\newcommand{\pcm}{\unit{cm}^{-1}} 
\newcommand{\pcmc}{\unit{cm}^{-3}}
\newcommand{\muJ}{\unit{\mu J}} 
\newcommand{\eV}{\unit{eV}}  
\newcommand{\Ohmcm}{\unit{\Omega\cdot cm}} 
\newcommand{\kel}{\unit{K}} 

\newcommand{\tisa}{\mbox{Ti:Al$_2$O$_3$}} 
\newcommand{\simx}{{\sim}}
\newcommand{\degrees}{\ensuremath{^\circ}}
\newcommand{\feff}{\ensuremath{\usr{f}{eff}}}

\newcommand{\water}{\ensuremath{\mathrm{H_2\mathrm{O}}}\xspace}
\newcommand{\cotwo}{\ensuremath{\mathrm{CO}_2}\xspace}

\newcommand{\nb}{\usr{n}{b}} 
\newcommand{\epsbr}{\usr{\epsilon}{br}} 
\newcommand{\depsr}{\usr{\Delta\epsilon}{r0}} 

\newcommand{\ffm}{Physikalisches Institut, J. W. Goethe-Universit\"at, Max-von-Laue-Strasse 1, 60438 Frankfurt am Main, Germany}

\begin{document}

% Use the \preprint command to place your local institutional report
% number in the upper righthand corner of the title page in preprint mode.
% Multiple \preprint commands are allowed.
% Use the 'preprintnumbers' class option to override journal defaults
% to display numbers if necessary
%\preprint{}

%Title of paper
\title{Ultrafast dynamic conductivity and scattering rate saturation 
of photoexcited charge carriers in silicon investigated with a midinfrared continuum probe}

% repeat the \author .. \affiliation  etc. as needed
% \email, \thanks, \homepage, \altaffiliation all apply to the current
% author. Explanatory text should go in the []'s, actual e-mail
% address or url should go in the {}'s for \email and \homepage.
% Please use the appropriate macro foreach each type of information

% \affiliation command applies to all authors since the last
% \affiliation command. The \affiliation command should follow the
% other information
% \affiliation can be followed by \email, \homepage, \thanks as well.
\author{Fanqi Meng}
\author{Mark D. Thomson}
%\author{Hartmut G. Roskos}
%\email[]{Your e-mail address}
%\homepage[]{Your web page}
%\thanks{}
%\altaffiliation{}
\affiliation{\ffm}

\author{Bo E. Sernelius}
\affiliation{
Division of Theory and Modeling, Department of Physics, Chemistry and Biology, Link\"{o}ping University,
58183 Link\"{o}ping, Sweden
}

\author{Michael J\"{o}rger}
\affiliation{Bruker Optik GmbH, Rudolf-Plank-Strasse 27  
76275 Ettlingen, Germany}

\author{Hartmut G. Roskos}
\email[Corresponding author: ]{roskos@physik.uni-frankfurt.de}
\affiliation{\ffm}

%Collaboration name if desired (requires use of superscriptaddress
%option in \documentclass). \noaffiliation is required (may also be
%used with the \author command).
%\collaboration can be followed by \email, \homepage, \thanks as well.
%\collaboration{}
%\noaffiliation

\date{\today}

\begin{abstract}
We employ ultra-broadband terahertz-midinfrared probe pulses to characterize the optical response of photoinduced charge-carrier plasmas in high-resistivity silicon
%, generated with $1.6\eV$ photoexcitation 
in a reflection geometry, over a wide range of excitation densities ($10^{15}-10^{19}\pcmc$) at room temperature.  In contrast to conventional terahertz spectroscopy studies, this enables one to directly cover the frequency range encompassing the resultant plasma frequencies.
The intensity reflection spectra of the thermalized plasma, measured using sum-frequency (up-conversion) detection of the probe pulses, can be modeled well by a standard Drude model with a density-dependent momentum scattering time of $\simx 200\fsec$ at low densities, reaching $\simx 20\fsec$ for 
densities of $\simx 10^{19}\pcmc$, where the increase of the scattering rate saturates.
This behavior can be reproduced well with theoretical results based on the generalized Drude approach for the electron-hole scattering rate, where the saturation occurs due to phase-space restrictions as the plasma becomes degenerate.
We also study the initial sub-picosecond temporal development of the Drude response, and discuss the observed rise in the scattering time in terms of initial charge-carrier relaxation, as well as the optical response of the photoexcited sample as predicted by finite-difference time-domain simulations.
\end{abstract}

% insert suggested PACS numbers in braces on next line
\pacs{}
% 42.65.Re 	Ultrafast processes; optical pulse generation and pulse compression
% 78.47.J- 	Ultrafast spectroscopy (<1 psec)
% 78.20.Bh 	Theory, models, and numerical simulation 
% 42.72.Ai 	Infrared sources

% insert suggested keywords - APS authors don't need to do this
%\keywords{}

%\maketitle must follow title, authors, abstract, \pacs, and \keywords
\maketitle

% body of paper here - Use proper section commands
% References should be done using the \cite, \ref, and \label commands

\section{Introduction}
The dynamics of charge carriers in silicon has been studied intensely over previous decades,
due to its importance in (opto-)electronic applications.
Spectroscopy provides an important tool for fundamental investigations, as it allows one to elucidate the intra-band charge-carrier dynamics via the frequency-dependent conductivity spectrum.
For carriers in doped Si, this response has been characterized over a very broad range of n-/p-type dopant carrier concentration ($10^{15}-10^{20}\pcmc$), employing measurements spanning the terahertz (THz) \cite{Exter90} and infrared \cite{Soref87} frequency ranges, where a Drude-type spectral behavior is generally observed, with modifications due to, e.g., inter-valence band transitions and energy-dependent carrier relaxation rates.
A number of reports have also been devoted to the study of photoexcited electron-hole (e-h) plasmas, not only because of their relevance for applications (optoelectronics \cite{Mnat05}, laser micromachining \cite{Lora00}, extreme surface nonlinear optics \cite{Baff14}) but also because they provide the possibility to study ultrafast relaxation/scattering processes vs. excitation density $\Nex$ and energy $h\nuex$.
Such pump-probe studies typically use inter-band excitation (UV--near-IR) and probe the spectral response in the THz \cite{Ziel1996,Hend2007}, mid-infrared (MIR) \cite{Naga2002} or optical \cite{Shank83,Hulin1984,Tint1995,Tint2000} range, depending on $\Nex$ (and hence the Drude plasma frequency $\nup$).
Other studies have estimated the time scales directly from time-domain data
using an ultra-short optical probe pulse \cite{Sabb2002} or photoelectron detection \cite{Ichi2009}.
Among these reports, various approaches are used to estimate the scattering time $\taus=\Gamma^{-1}$ from the data.  
In reviewing the range of reported data (presented in this paper), we found that there is some disparity in the values and their dependence on $\Nex$, and at high excitation densities ($\Nex\gtrsim 10^{20}\pcmc$) there still appears to be contention about the order of magnitude for $\taus$.  
For certain experiments (e.g. those involving the charge-carrier plasma as a moving front for Doppler up-conversion \cite{Thomson13,Meng14a}), the precise magnitude of $\taus$ sensitively affects the degree of absorption loss for probe frequencies (especially about $\nup$), and hence a quantitative determination $\taus$ vs. $\Nex$ is necessary.  
While the role of the various scattering mechanisms has been investigated theoretically \cite{Comb1987-1,Comb1987-2,Sern1987,Sern1987-2,Sern1991}, a detailed comparison between theory and experiment is also lacking.

In the present paper, we apply pump-probe spectroscopy on Si(100) at $T=293\kel$ in a reflection geometry, using both ultra-broadband MIR-probe and conventional THz-probe pulses, to extract estimates of $\taus$ (following initial energy relaxation) over the range $\Nex=3\cdot 10^{15}-2\cdot 10^{19}\pcmc$, by applying a standard Drude model to fit the intensity spectra (and accounting for the longitudinal excitation density profile in the analysis).
A saturation in the increase of the scattering rate is observed for $\Nex\simx 10^{19}\pcmc$, with $\taus\simx 20\fsec$.
This behavior can be reproduced well by theory which includes the e-h scattering rate (whose variation dominates the density dependence in this range) via a generalized Drude approach \cite{Sern1987,Sern1991}.  This supports the assertion that the saturation is due to the onset of phase-space restrictions as the plasma becomes degenerate (and the density of vacant final-states reduces).
While the present time resolution is limited to a few hundred femtoseconds, we also investigate the sub-picosecond evolution of the plasma response.  A comparison to model results from one-dimentional (1D) finite-difference time domain (FDTD) simulations shows that while one should be wary of artifacts in the fitted values of $\taus$ during the initial rise of the signal, the data indeed indicate that the scattering rate relaxes to a steady state-value during the first $1\psec$ following excitation.

\section{Experimental details}

\begin{figure}%
\includegraphics[width=0.95\columnwidth]{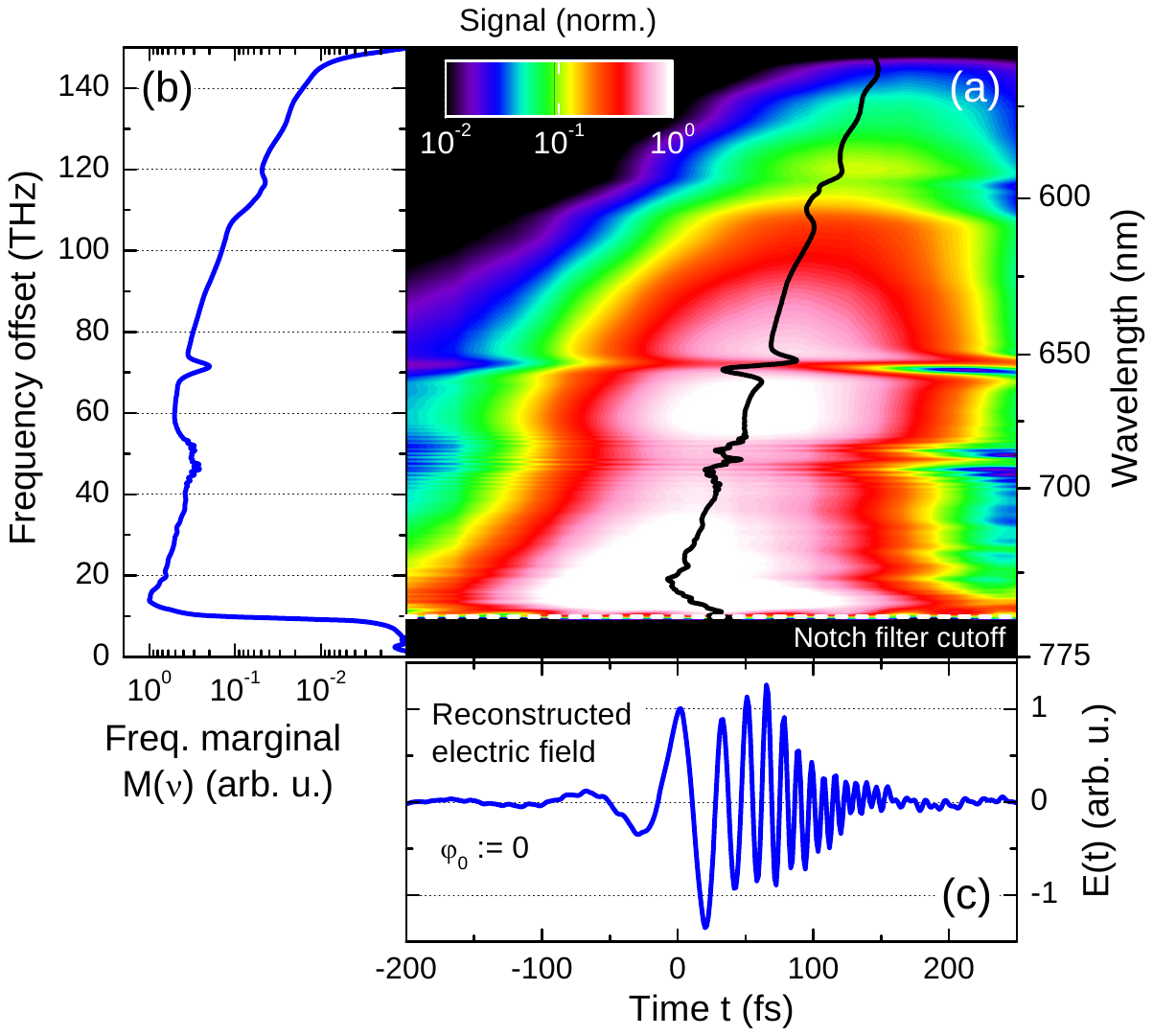}%
\caption{
(a) Spectrogram of ultra-broadband THz-MIR continuum probe pulse (black curve indicates temporal first-moment vs. frequency, $\tmu$). (b) Corresponding 
spectral marginal $\Mnu$. (c) Estimate of temporal electric field profile $E(t)$, obtained using $\Mnu$ and $\tmu$ (assuming a carrier-envelope phase of $\varphi_0=0$, see text).}%
\label{fig:refsgram}%
\end{figure}

The experimental system is based on a 1-kHz \tisa-amplifier laser (Clark-MXR, CPA-2101, $\lambda_0=775\nm$, $\Tfwhm=150\fsec$).
%, which is used to provide the sample excitation pulses (energy up to $200\muJ$) and optical detection gate pulses.
For the ultra-broadband THz-MIR probe pulses, we employ the emission from a two-color air-plasma (as reported previously \cite{Thomson2010,Blank13}), whose pump beam (energy $380\muJ$) is obtained by spectral broadening of the Chirped Pulse Application (CPA) laser pulses in an Ar-filled (2.5~bars) hollow-core fiber and subsequent recompression with a set of negative-dispersion mirrors. The pump beam was focused to generate the plasma with a planoconvex lens ($f=200\mm$) via a $150\mum$-thick $\beta$-BBO crystal ($32\degrees$-cut) to provide the second-harmonic beam.
The ultra-broadband emission had a typical pulse energy of $100\unit{nJ}$  (directly after the plasma, as inferred from measurements with a calibrated pyroelectric detector) and spectral coverage extending to $\simx 150\thz$ ($5000\pcm$, $0.62\eV$).
This beam is collimated by an off-axis paraboloidal mirror (OAPM, effective focal length $\feff=152.4\mm$) after passing a $450\mum$-thick Si wafer (to discard the optical pump beams).
The optical-pump THz-MIR-probe reflection measurements were performed by deflecting the probe beam with a $450\mum$-thick Si beamsplitter to the sample, where it was focused by an OAPM ($\feff=101.6\mm$) at normal incidence, reflected, and transmitted through the Si beamsplitter to the detection stage.  The CPA sample-pump beam (with pulse energy up to $\simx 50\muJ$) was brought to a diameter of $0.6\mm$ (FWHM) on the sample, using a non-collinear geometry.
%(see inset in Fig~\ref{fig:excdep})
For detection of the THz-MIR probe pulses, we employed sum-frequency (SF) generation (up-conversion) \cite{Thomson13b} in a $500\mum$-thick $\mval{110}$-cut ZnTe crystal with 150-fs optical detection pulses, whereby the THz-MIR beam is collinearly focused with an OAPM ($\feff=101.6\mm$).  Following the SF crystal, a notch filter is used to discard the input optical light and the SF spectrum is measured by coupling free-space into a miniature spectrometer with a cooled CCD sensor (Ocean Optics QE65).  Integration times of 50-100~ms are sufficient to acquire a SF spectrum for each value of detection delay $t$.
A measured reference spectrogram $S(\nu,t)$ of these pulses (where $\nu$ is the frequency offset from $\nu_0=c/\lambda_0$) is shown in Fig.\ref{fig:refsgram}(a). Here the beam was sent directly to the detection (and hence represents that which is incident on the sample for the measurements below). In order to estimate the intensity spectrum $I(\nu)$, we also calculate the frequency marginal $M(\nu)={\int}\ud t \, S(\nu,t)$ as shown in Fig.~\ref{fig:refsgram}(b).  
Note that the spectrogram (and hence $M(\nu)$) is affected both by the phase-matching response (which for such thick crystals varies as $P(\nu,\nu_0)\simx 1/\Delta k(\nu,\nu_0)$ \cite{Thomson13b}), as well as a convolution with the spectral optical detection spectrum (with width $\Delta\nu=3.8\thz$ here).  Nevertheless, for the calculation of reflectivity spectra, $M(\nu)$ can be used with reasonable accuracy, as long as the spectral resolution is included in subsequent analysis.
One can see that the measured signal extends from the cut-off of the notch filter ($\simx 10\thz$) up to $\simx 150\thz$. The modulation in the spectrum is due to the absorption lines from ambient \water ($\simx 50$ and $110\thz$) and \cotwo ($\simx 70\thz$) in the beam path.  While the temporal width of the spectrogram is dictated by the optical detection pulses, one can still extract an estimate of the group delay $\usr{T}{g}(\nu)$ via the temporal first moment vs. frequency, i.e. $\tmu={\int}\ud t\,t\cdot S(\nu,t)/{\int}\ud t\,S(\nu,t)$, which is also included in Fig.~\ref{fig:refsgram}(a) (black line), and indicates a pulse duration of $\simx 100\fsec$, which is due to primarily to the dispersion of the $450\mum$-Si wafer.  While only approximate, to give an impression of the time-domain pulse, in Fig.~\ref{fig:refsgram}(c) we plot a reconstructed field profile using the approximation $E(\nu)\simx\sqrt{M(\nu)}e^{-i{\int}_0^\nu\ud \nu\, \tmu}$ (note that SFG detection can only provide the relative spectral phase, and we assume an arbitrary value of $\varphi_0=0$ for the carrier-envelope phase here to illustrate the temporal chirp of the carrier wave).

The purity of the high-resistivity Si samples used (float zone, $(100)$, thickness of $525\mum$, $\rho>3000\Ohmcm$, Crystec GmbH) was characterized by
two different spectroscopic methods. 
Firstly, shallow group III and V impurities were analyzed by low-temperature near-IR photoluminescence at $T=4.2\kel$ using a Bruker Vertex80 FTIR spectrometer with a photoluminescence module and liquid-helium immersion cryostat. 
Quantification of impurities was carried out according to the SEMI MF1389 standard \cite{Semi} using a suitable set of calibration samples. 
Excitation with two different wavelengths ($\lambda_{1}= 532\nm$, absorption depth $\absex\simx 1 \mum$; $\lambda_{2}= 976\nm$, $\absex\simx 70 \mum$) gave similar results, indicating no significant gradient of the shallow impurity concentration. 
Secondly, substitutional carbon and interstitial oxygen content were analyzed according to the SEMI MF1391 and SEMI MF1188 standards with room-temperature MIR absorption spectroscopy using a Bruker Vertex70 FTIR spectrometer.
The concentrations (upper bounds) of III/V dopants and C/O are listed in Table~\ref{tab:si}, which indicate that the sample purity is not corrupted e.g. by significant compensation doping, compared to the excitation densities used here.

%\bgroup
%\def\arraystretch{1.5}%
\begin{table}%
\def \colwidA{0.06\columnwidth}
\def \colwidB{0.24\columnwidth}
\begin{tabular}{
L{\colwidA} 
L{\colwidB} 
L{\colwidA} 
L{\colwidB} 
L{\colwidA} 
L{\colwidB}
}
\hline
\rule{0pt}{3ex}    
B & $\simx 2.9\cdot 10^{11}$ & 
P & $\simx 7.9\cdot 10^{11}$ & 
C & $<5\cdot 10^{16}$  \tabularnewline 
Al & $<5\cdot 10^{10}$ & 
As & $<5\cdot 10^{10}$ &
O & $<5\cdot 10^{16}$   \tabularnewline
& & Sb & $<5\cdot 10^{10}$ & &   \tabularnewline
\hline
\end{tabular}
\caption{Determined elemental impurity concentrations (in $\pcmc$) in Si samples from low-temperature photoluminescence (4 K) and mid-IR absorption (300 K) measurements.}
\label{tab:si}%
\end{table}%
%\egroup

\section{Results}
\subsection{Excitation density dependence}\label{sec:excdep}

\begin{figure}%
\includegraphics[width=0.9\columnwidth]{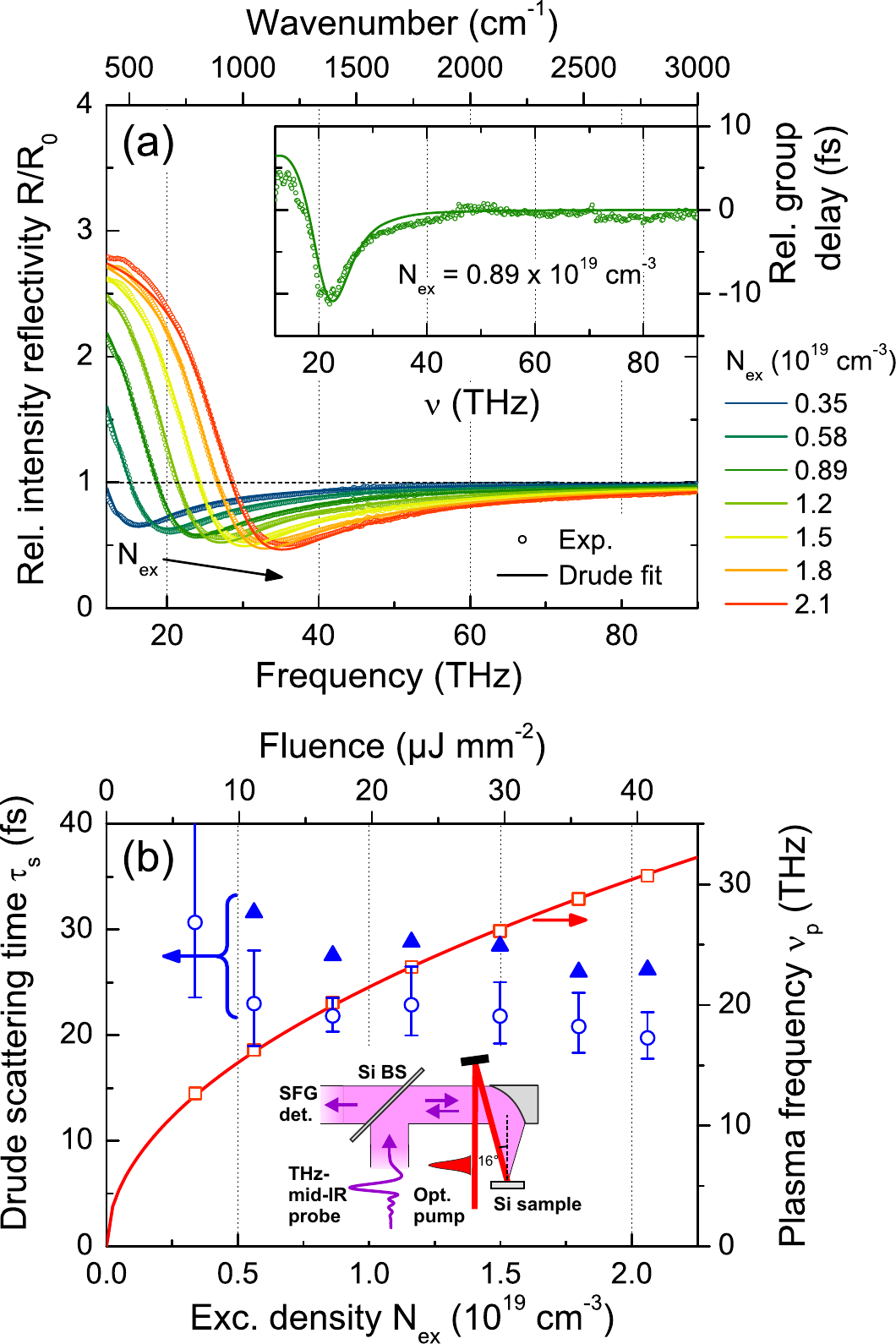}%
\caption{
(a) Experimental intensity reflection spectra $R(\nu)=M(\nu)/M_0(\nu)$ for different values of optically excited plasma density $\Nex$ ($\lamx=775\nm$) and a pump-probe delay of $\tau=1\psec$, and fitted model curves assuming a standard Drude dispersion model and accounting for the finite-depth exponential excitation density profile (with $\absex=125\pcm$).  The inset shows the corresponding group delay $\Delta\tmu$ and model curve using parameters from fitting intensity reflection.
(b) Corresponding parameters $\taus$ ($\medcircle$), $\nup$ ($\medsquare$), from model fits in (a) vs. $\Nex$. 
Also included are fitted values of $\taus$ ($\blacktriangle$) when using only the SF spectra at $t=0$, for later discussion in connection with the data in Fig.~\ref{fig:tausgram}.  The inset depicts schematic of experimental reflection geometry.
}%
\label{fig:excdep}%
\end{figure}

Figure ~\ref{fig:excdep}(a) shows the measured intensity reflection spectra $R(\nu,\tau)/R_0(\nu)$ (i.e., relative to the reflection $R_0$ of the unpumped sample, which is a flat curve with $R_0=0.30$) at a pump-probe delay of $\tau=1\psec$ and a range of front-face ($z=0$) excitation densities $\Nex=0.35 \mhyp 2.1\cdot 10^{19}\pcmc$.  

Each reflection spectrum was fitted assuming a standard Drude model for the plasma conductivity
$\sigma(\omega,z)=e^2\Nex(z)/\meff(1+i\omega\taus)^{-1}$, 
where $\meff=0.15 m_0$ is taken for the effective reduced e-h mass, and $\Nex(z)=\Nex e^{-\absex z}$ accounts for the $z$-dependent longitudinal excitation.
It is well established for the reflectivity of such an excitation profile (and probe wavelengths $\lambda=c/\nu\gtrsim \absex^{-1}$), that one should not use simply the Fresnel field reflection coefficient $\usr{r}{F}$ (i.e. based on the front-face density) for quantitative analysis, rather one must account for the distributed reflection of (and losses within) the excitation profile \cite{Vine1984,Ma89}.
The solution for the total field reflectivity at normal incidence can still be expressed in closed-form, which upon inspection of the formulas in Ref.~\onlinecite{Vine1984}, can be expressed as $r=(1-n')/(1+n')$, where 
$n'=\nb+\sqrt{\depsr}X_\beta(\xi)$,
$\nb=\sqrt{\epsbr}$ is the background (unpumped) refractive index, 
$\depsr(\omega)=-i\sigma(\omega,0)/\epsz\omega$ is the front-face photoinduced change in the complex permittivity, and
$X_\beta(\xi)=I_{\beta+1}(\xi)/I_{\beta}(\xi)$ is the ratio of modified Bessel functions with 
$\beta=2i\omega\nb(c\absex)^{-1}$ and
$\xi=2i\omega\sqrt{\depsr}(c\absex)^{-1}$.  During the fitting, we convolved the raw model spectra with the spectral detection response (width $\Delta\nu$ given above).

The fitted model curves are included in Fig.~\ref{fig:excdep}(a) and are generally in close agreement with the experimental data. The corresponding fit parameters ($\taus$ and $\nup=\omp/2\pi$ where 
%$\omp=e^2\Nex/\epsb\epsz\meff$)
$\omp^2=\sigma(0,0)/\epsb\epsz$)
are shown in Fig.~\ref{fig:excdep}(b). 
We firstly note that the values of $\nup$ correspond closely to a $\sqrt{\Nex}$-dependence (solid curve), which supports the idea that $\Nex\propto F$ (where $F$ is the excitation fluence) and shows that higher-order excitation effects (e.g. two-photon absorption \cite{Tint2000}) do not play a role in these measurements.
The values of the Drude scattering time $\taus$ exhibit a moderate trend 
of increasing scattering rate $\Gamma=\taus^{-1}$ with $\Nex$, ranging from $\taus=30.7\fsec$ ($\Nex=0.35\cdot 10^{19} \pcmc$) to
$\taus=19.7\fsec$ ($\Nex=2.1\cdot 10^{19}\pcmc$), as discussed below.  We note that the fit residuals in Fig.~\ref{fig:excdep}(a) typically deviated beyond the random noise levels in the data, i.e. a small systematic disagreement exists for each dataset, precluding a standard fit-parameter error analysis \cite{numrecipes}.
In order to obtain confidence intervals for the fitted values of $\taus$, we analyzed the rms-misfit $O=\sqrt{\sum(R-\usr{R}{mod})^2}$ in the (\nup,\taus)-neighborhood of the fitted parameters and give here the bounds for $\taus$ for the elliptical region where $O=\sqrt{2}\usr{O}{min}$.
Note that while we fitted here only the intensity curves $R(\nu)=|r(\nu)|^2$, for the field reflectivity response $r(\nu)$ one also expects a spectral phase corresponding to a small but measurable group delay $\usr{\Delta T}{g}(\nu)$.
This can also be estimated from the spectrograms via $\Delta\tmu=\tmu-\tmuz$, as described above. In the inset of Fig.~\ref{fig:excdep}(a) we plot the experimental $\Delta\tmu$ data and the model curve obtained using the intensity-fit parameters for the case $\Nex=0.89\cdot 10^{19}\pcmc$, which are seen to be in good agreement and further demonstrates that the experimental data correspond well to the Drude/reflection model used.

%\subsubsection{Comparison with literature data}

\begin{figure}%
\includegraphics[width=0.95\columnwidth]{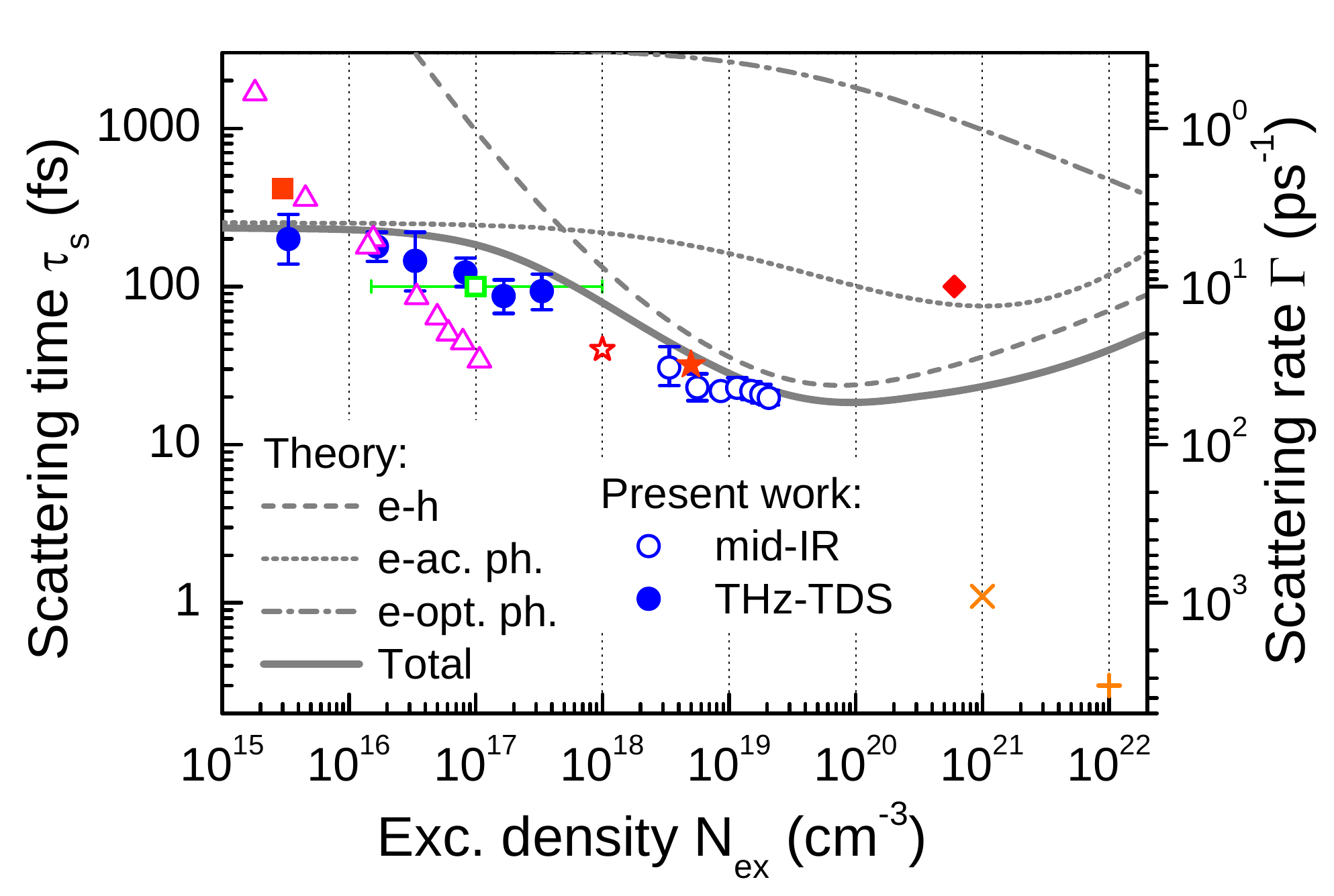}%
\caption{
Experimental and theoretical (Drude) scattering time vs. $\Nex$.  Values from present work ($\lamx=775\nm$) using THz-TDS (\ding{108}, see text) and mid-IR continuum probe with SFG-detection  ($\medcircle$, data from 
\ref{fig:excdep}(b)).
Literature sources (incl. $\lamx$, at room temperature unless otherwise stated): 
%$\filledmedsquare$: $760\nm$ \cite{Ziel1996};
%$\triangle$: $400\nm$, $\blacktriangle$: $266\nm$ \cite{Hend2007};
%$\medsquare$: $800\nm$ \cite{Tsub2012};
%$\medstar$: $610\nm$ \cite{Ichi2009};
%\ding{72}: $800\nm$ \cite{Sabb2002};
%\ding{117}: $530/1060\nm$ \cite{Drie1987};
%\ding{53}: $625\nm$ \cite{Tint2000};
%\ding{58}: $620\nm$ \cite{Hulin1984}. 
Ref.~\onlinecite{Ziel1996} ($\filledmedsquare$, $760\nm$);
%Ref.~\onlinecite{Hend2007} ($T=30$~K, $\triangle$, $400\nm$; $\blacktriangle$, %$266\nm$);
Ref.~\onlinecite{Hend2007} ($\triangle$, $400\nm$, $T=30\kel$);
Ref.~\onlinecite{Tsub2012} ($\medsquare$, $800\nm$);
Ref.~\onlinecite{Ichi2009} ($\medstar$, $610\nm$);
Ref.~\onlinecite{Sabb2002} (\ding{72}, $800\nm$);
Ref.~\onlinecite{Drie1987} (\ding{117}, $530\nm$);
Ref.~\onlinecite{Tint2000} (\ding{53}, $625\nm$);
Ref.~\onlinecite{Hulin1984} (\ding{58}, $620\nm$). 
Curves correspond to theoretical calculations of the total scattering rate (solid) from e-h (dashed) \cite{Sern1991}, acoustic- (dotted) and optical-phonon (dash-dot) scattering \cite{Comb1987-1}, respectively (see text).
}%
\label{fig:excdeplit}%
\end{figure}

In order to put these measured values of $\taus$ in perspective, in Fig.~\ref{fig:excdeplit} we plot the values along with various previously reported estimates from the literature (as indicated in caption) vs. $\Nex$.  In order to provide additional data for lower $\Nex$, we also performed a set of optical-pump--THz-probe reflection measurements using a conventional THz time-domain spectroscopy (TDS) system covering the range $0.1-3\thz$. The obtained values of $\taus$ (for $\Nex=3.3\cdot 10^{15}-3.3\cdot 10^{17}\pcmc$ at $\tau=+5\psec$) are also included in  Fig.~\ref{fig:excdeplit} (where we applied the same model as above, except here fitting the complex reflectivity data $r(\nu)$).  
The literature data were obtained using various techniques, some of which do not directly probe precisely the Drude scattering, i.e., current damping, rate, including reports in which the values are extracted directly from analysis and those in which estimates are cited that are found to be consistent e.g. with the results of auxiliary simulations (and as such may only indicate an approximate value of $\taus$).

For low excitation densities, our results give a value of $\taus=200\pm 75 \fsec$, which is close to cited values for the e-ph scattering time $\Gameph^{-1}$ at $T=300\kel$ of $240 \mhyp 260\fsec$ \cite{Sjod1998,Sabb2002,Ichi2009}.
A previous pump-probe THz-TDS study ($\filledmedsquare$, $\Nex=3 \cdot 10^{15}\pcmc$ \cite{Ziel1996}) estimated a value of $\taus=330-500\fsec$ ($\Gamma/2\pi=2\mhyp 3\thz$), although this report concentrated on simulating the response during the initial dynamics (discussed further below) and accounting for the finite pulse duration of the THz probe pulse, as opposed to a rigorous analysis of the conductivity spectra of a quasi-static plasma at sufficiently large delay $\tau$.
A more recent study using THz-TDS \cite{Hend2007} at $T=30\kel$ with  ($\lamx=400\nm$, $\triangle$) fitted the complex conductivity spectra $\sigma(\nu)$ for $\tau=+10\psec$ as a function of $\Nex$. 
Given the low temperature, the higher $\taus$ values ($\simx 2\psec$) at low densities are reasonable, as e-ph scattering is strongly suppressed \cite{Comb1987-3}). However, the observed trend with $\Nex$ differs significantly from our results, as discussed further below. 
Our data are consistent with the value of $\taus=100\fsec$ employed in Ref.~\onlinecite{Tsub2012} ($\medsquare$) used to simulate experiments involving the reflection from a counter-propagating plasma boundary in the range 
$\Nex=1.5\cdot 10^{16} \mhyp 10^{18}\pcmc$ (although no error margin for this value was given).

In the range $\Nex=10^{18}-10^{19}\pcmc$, two reports provide estimates of carrier scattering time which are reasonably consistent with our data, although they do not specifically probe the current relaxation rate.
In Ref.~\onlinecite{Ichi2009} ($\medstar$, $\Nex\simx 10^{18}\pcmc$), optical two-photon photoemission measurements were used to deduce an initial time scale for momentum relaxation of $\taus=40\fsec$, whereas in
Ref.~\onlinecite{Sabb2002} (\ding{72}, $\Nex=5.5\cdot 10^{18}\pcmc$) a time scale of $\taus=32\pm 5\fsec$ was determined from the decay of the coherent four-wave-mixing signal.  In both of these reports, this time scale accounts specifically for the initial (elastic) momentum reorientation and redistribution of the carriers, while the energy relaxation time (due to e-ph scattering) was determined through additional measurements to be $\taueph\simx 250\fsec$. Hence these studies include the e-h and e-e scattering contributions directly, the latter of which is only expected to contribute to the Drude current damping via the conduction-band anisotropy \cite{Sern1991}. 
We note that another study \cite{Naga2002} of MIR reflectivity spectra (similar to the MIR measurements here only with a point-wise probing with tunable source) for $\Nex\simx 10^{19}-10^{20}\pcmc$ used a nominal value of $\taus=100\fsec$ to provide a set of additional simulated curves.  However, as no attempt was made to extract $\taus$ from the experimental data, we omit this value in Fig.~\ref{fig:excdeplit}.

The remaining data at higher excitation density $\Nex>10^{20}\pcmc$ all involved using a single probe wavelength and varying $\Nex$ for determining the Drude response.
The value of $\taus=100\fsec$ in Ref.~\onlinecite{Drie1987} (\ding{117}, $\Nex\simx 6\cdot 10^{20}\pcmc$), which tends to deviate significantly from the overall trend of the data, was cited in connection with detailed simulations of the delay-dependent reflectivity of a 20-ps probe pulse ($\lambda=2.8\mum$), although no estimate of the error was provided and perhaps should be taken to indicate only the order of magnitude of $\taus$.
The extreme excitation densities (and hence $\nup$ values) of Ref.~\onlinecite{Tint2000} (\ding{53}, $\Nex\simx 10^{21}\pcmc$) and 
Ref.~\onlinecite{Hulin1984} (\ding{58}, $\Nex\simx 10^{22}\pcmc$) required the use of a visible-range probe, with estimates $\taus \simx 1.1\fsec$ and $\simx 0.3\fsec$ obtained by fitting the Drude-type reflection curves, respectively. While the fitted curves around $\nu=\nup$ in those reports are in reasonable agreement with the experimental data (although in both reports there are deviations on the high-$\Nex$ side), it is not clear whether such an approach can be applied robustly to extract $\taus$.
In simulations, we found that the extracted value of $\taus$ from reflectivity spectra is particularly sensitive to any distortion of the curves, as it is essentially the curvature of $R(\nu)$ about $\nup$ which allows its determination.  Given that a Drude analysis of data with a single probe wavelength necessitates the variation of $\Nex$ (and hence $\taus$), distortions of the measured $R(\Nex)$ curves are to be expected (compared to spectra $R(\nu)$ at constant $\Nex$), and we exercise caution in considering these values.  
We note that in both reports, it is the initial sub-picosecond carrier plasma response that is probed -- i.e. in Ref.~\cite{Tint2000}  $R(\Nex)$ is measured with a pump-probe delay of $\tau=150\fsec$ with 100-fs pulses, while in Ref.~\cite{Hulin1984} the self-induced reflection changes are measured with 90-fs pulses.
However, in a preceding experimental report \cite{Tint1995}, the authors of Ref.~\onlinecite{Tint2000} and colleagues extracted $\taus$ from their data vs. pump-probe delay (for a fixed $\Nex\simx 10^{22}\pcmc$), asserting that $\taus\approx 0.5\fsec$ for the whole measured delay range out to $\tau=0.8\psec$. Note that for those measurements, a single probe wavelength was used with an analysis of p- and s-reflectivity at oblique incidence to extract the complex refractive index and hence Drude parameters, as opposed to analysis of $R(\Nex)$-curves.
To summarize, while some concern has been raised about the plausibility of these values \cite{Sern1991}, the assertion in the literature that scattering times $\taus\simx 1\fsec$ are obtained at these high densities (even for time scales $\gtrsim 100\fsec$) has not yet been refuted.

\subsection{Theoretical predictions}

We now turn to theoretical predictions for the observed dependence on $\Nex$.
Qualitatively, the e-phonon scattering rate $\Gameph$ is expected to vary only weakly with $\Nex$ for densities below $\simx 10^{20}\pcmc$ \cite{Comb1987-3}, whereas (bimolecular) \mbox{e-h} scattering should obey $\Gameh\propto\Nex$ for low densities. While like-charge scattering does not contribute to current damping for isotropic bands, for the anisotropic X-valleys in Si \mbox{e-e}-scattering can also contribute, although in the low-frequency limit this is predicted to have only a very minor influence on the measured scattering time \cite{Sern1991}.
The situation for increasing $\Nex$ was closely examined theoretically \cite{Comb1987-1,Sern1987,Sern1991} following the sub-femtosecond experimental value found in Ref.~\onlinecite{Hulin1984}.
Using different approaches, they all predict that as the plasma becomes degenerate ($\Nex\gtrsim \Nexthr$, where $\Nexthr=(2\mDOS kT/\hbar^2)^{3/2}/3\pi^2$, \cite{Comb1987-1}) $\Gameh$ should reach a maximum value and then begin to decrease with $\Nex$, due to the reduction of available final states for scattering (phase-space restrictions in the quantum limit \cite{Sern1987}).  
For electrons in the conduction band at $T=300\kel$ one has $\Nexthr=3.4\cdot 10^{18} \pcmc$, using the value $\mDOS=0.32\cdot\mel$ for the (valley-degeneracy-scaled) DOS mass. 
Although the effective temperature $T$ can be expected to rise with excitation fluence due to the excess excitation photon energy (for $h\nuex>\usr{E}{g}=1.12\eV$), at $T=300\kel$ this amounts to only some $10\kel$ for $h\nuex=1.6\eV$.
Hence one expects that the e-h scattering dominates the density dependence of $\Nex$, and that a saturation of the $\Gamma\propto \Nex$ dependence should be observed for $\Nex\simx 10^{18} \mhyp 10^{19}\pcmc$, as is indeed indicated in our experimental results.

We carried out calculations of $\Gameh$ vs. $\Nex$ using the theoretical treatment in Ref.~\onlinecite{Sern1991}, i.e. based on a generalized Drude model, where the e-h scattering is calculated from integration over the appropriate expression of the complex polarizability functions $\usr{\alpha}{e,h}(\mathbf{q},\omega)$ to yield the relevant coefficients of mutual friction, which are then used to evaluate the complex conductivity $\sigma(\nu)$.  Here we assume isotropic bands for simplicity (with effective masses $\usr{m}{de}=0.32\cdot \mel$ and $\usr{m}{h}=0.52\cdot \mel$) and calculate the low-frequency limit for $\Gameh$ (which is a reasonable approximation for the spectral range used in our experiments) using the finite-temperature ($T=300\kel$) expression for 
$\operatorname{Im}\{{\usr{\alpha}{e,h}}\}$ 
and zero-temperature expression for 
$\operatorname{Re}\{{\usr{\alpha}{e,h}}\}$.
The calculated rates $\Gameh$ vs. $\Nex$ are shown in Fig.~\ref{fig:excdeplit} (dashed curve), which indeed exhibit a saturation of the $\Gameh\propto \Nex$ dependence in range mentioned above, reaching a minimum with $\taus\sim 20\fsec$ at $\Nex=8\cdot 10^{19}\pcmc$ above which the e-h scattering rate begins to drop again. 

In order to include the contribution of e-ph scattering, we employed the theoretical expressions in Ref.~\onlinecite{Comb1987-1} (Eqs. 7 and 8 for the optical and acoustic phonon scattering, respectively). The calculated curves are also shown in Fig.~\ref{fig:excdeplit} (using the following values:
LO optical phonon deformation potential $D_1=6\cdot 10^8 \eV \units{cm}{-1}$ \cite{pop2004}; 
optical phonon energy $\hbar\usr{\Omega}{oph}=64.4\unit{meV}$ \cite{dargys94};
acoustic deformation potential $E_1=8.2\eV$ \cite{Fisch00};
sound velocity $\usr{v}{s}=9\cdot 10^{5} \unitd{m}{}{s}{-1}$).  
As expected, the acoustic phonon scattering rate $\usr{\Gamma}{e\mhyp aph}$ dominates in the low-density limit, with a value $\taus\approx 250\fsec$ reproducing experimental values well.  Based on this model, $\usr{\Gamma}{e\mhyp aph}$ increases gradually (with a $\Nex^{1/3}$-dependence due to an increase of the electron velocity) reaching a maximum at $\Nex\approx 10^{21}\pcmc$, upon which it decays with $\Nex^{-1}$ due to the additional onset of phase-space restrictions (as per $\Gameh$ above).  The contribution of optical phonon scattering is  marginal, increasing only at very high densities where electrons in the distribution increasingly reach energies $\simx \hbar\usr{\Omega}{oph}$.

The combined scattering rate is also shown in Fig.~\ref{fig:excdeplit} (solid curve), and is seen to reproduce the trend and magnitude of our experimental data well. To our knowledge, such a correspondence between experiment and theory for the Drude scattering time for carrier plasmas in undoped Si vs. $\Nex$ has not been demonstrated until now.
This correspondence has an important impact for judging the overall trend in the experimental data in Fig.~\ref{fig:excdeplit}.  On the basis of the experimental  data alone, one might consider that the very short reported values of $\taus$ for $\Nex\geq 10^{21}$ could be consistent with a $\Gamma\propto\Nex$-dependence extending to these high densities (as has been previously asserted \cite{Hulin1984}), with the apparent saturation of the values from our measurements for $\Nex\simx 10^{19}\pcmc$ possibly being due to unforeseen artifacts in the experimental method.
However, the consistency in the present results seems to rule this out, at least for charge-carrier plasmas after any carrier-carrier thermalization on a sub-100-fs time scale (see next section).  Clearly the treatment here of e-h and e-ph scattering does not predict scattering times of $\taus\simx 1\fsec$ at high density.
We also have difficulties to reconcile the data from Ref.~\onlinecite{Hend2007} (Fig.~\ref{fig:excdeplit}, $\triangle$). In considering the low temperature for those measurements, one expects the onset of plasma degeneracy at even lower densities than for our experiments.  Further experiments vs. $T$ and excitation energy are required to resolve this issue, where a broader probe spectral range including the plasma frequency is covered, as per our measurements here.

\subsection{Sub-picosecond relaxation dynamics}

In this last section, we present time-resolved measurements of the plasma response probed with the THz-MIR pulses. 
As will be shown, while certain aspects of the data would necessitate superior time-resolution than that available here, we can draw important conclusions regarding the analysis and interpretation of such measurements via comparison with simulations. 

\begin{figure}%
\includegraphics[width=0.85\columnwidth]{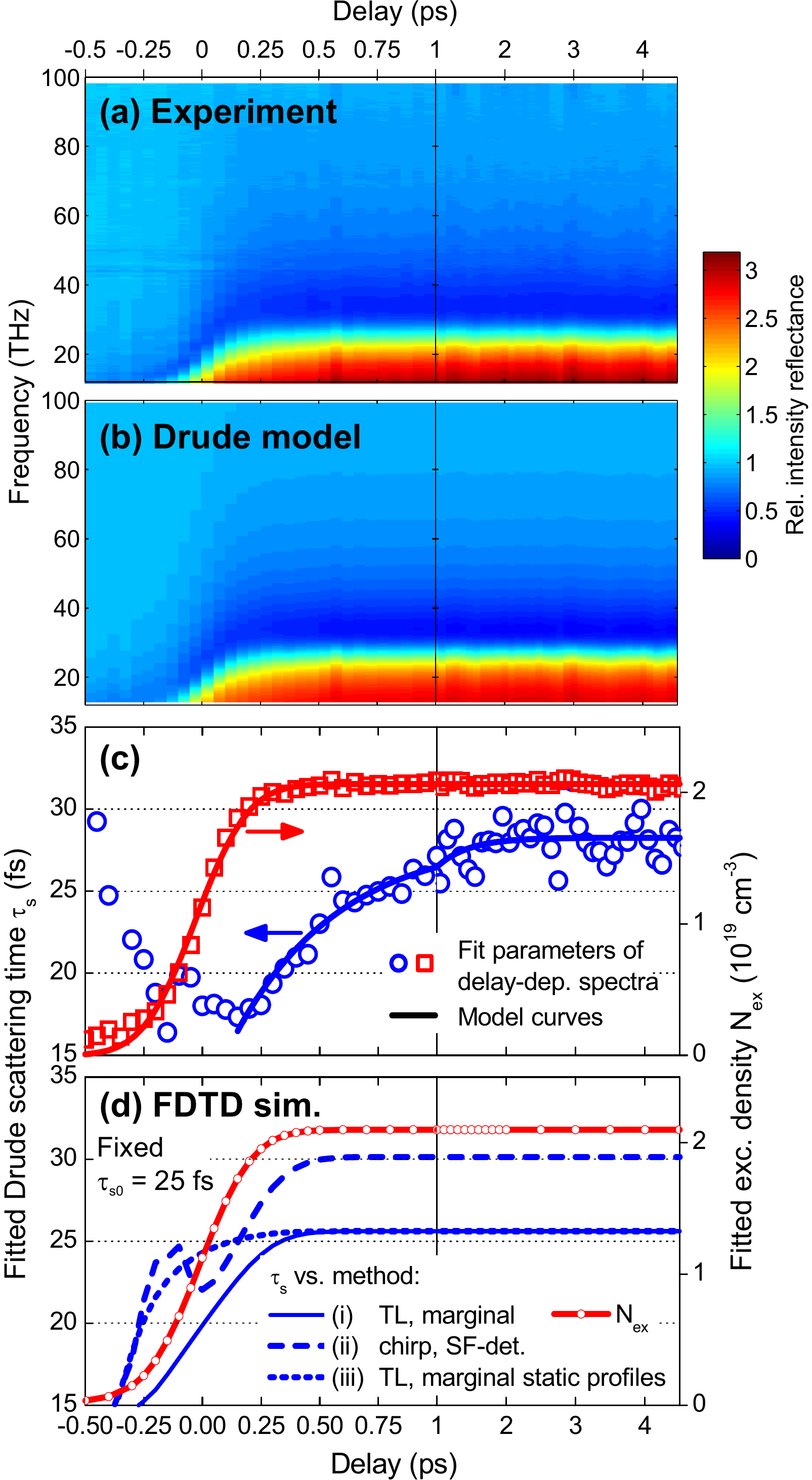}%
\caption{
(a) Experimental relative intensity reflection spectra vs. pump-probe delay $\tau$, estimated using SF spectra at fixed detection time $t_0$, i.e. $R'(\nu,\tau)=S(\nu,t_0;\tau)/S_0(\nu,t_0)$, where $S_0(\nu)$ is the reference spectrum before excitation (measured at sufficiently negative $\tau$). (b) Corresponding model spectra using Drude model fits for each value of $\tau$. (c) Fitted Drude parameters  $\taus$ ($\medcircle$), $\Nex$ ($\medsquare$) for model data in (b). (d) Corresponding analysis of FDTD simulation results with a constant value of $\tausz=25\fsec$, demonstrating artifact rise of fitted $\taus$ values around zero delay for three simulation batches: (i) transform-limited (TL) THz-MIR pulse with full-spectra (solid), 
(ii) including experimental THz-MIR pulse chirp and simulating SF detection at fixed delay (dashed), and (iii) with a TL probe pulse and a sequence of static, partially formed plasma profiles (short dash).}%
\label{fig:tausgram}%
\end{figure}

The time-resolved THz-MIR reflectivity spectra $R'(\nu,\tau)$ were measured in the same experimental geometry as the data in Fig.~\ref{fig:excdep}(a) with $\Nex=2.1\cdot 10^{19}\pcmc$, only here, in order to achieve practical measurement times we acquired each SF spectrum $S(\nu,t_0;\tau)$ at a fixed time $t=t_0$, as opposed to acquiring a full spectrogram to calculate the marginal (hence the measurements capture the spectral components within the 150-fs SF detection gate pulse).
The data are shown in Fig.~\ref{fig:tausgram}(a).  For each spectrum, we employed the same Drude model analysis as above, in order to yield $\Nex$ and $\taus$ as a function of $\tau$.
The model spectra and corresponding fit parameters are shown in Fig.~\ref{fig:tausgram}(b,c), respectively.  As can be seen, the data can be reproduced well over the full delay range with this model.  The kinetics of the fitted $\Nex$-data can be fit well assuming a Gaussian response (i.e. error-function dependence) with a FWHM width of $\usr{T}{r}=420\fsec$ (due to the combination of the pump-probe correlation function and non-collinear geometry).  
The corresponding kinetics of the $\taus$-values show a rise from 
$\taus\simx 17\fsec$ around $\tau=0$ to reach a value of $\taus=28\fsec$ for a delay of some ps.  The continuing rise of the $\taus$-curve compared to that of $\Nex$ clearly indicates a subsequent evolution of the plasma scattering rate following excitation. This trend is not consistent with that expected simply for an increasing plasma density during the excitation pulse, and is addressed in detail below.
A comparison of the value for $\tau=+1\psec$ ($\taus\approx 26\fsec$) with the corresponding data in Fig.~\ref{fig:excdep}(b) ($\taus\approx 19.7\fsec$) reveals an apparent discrepancy (both sets of data were acquired in the same measurement run).  As the only difference in the two measurements is that in one case the full SF marginal $M(\nu)$ is acquired vs. the SF spectra at fixed $t=t_0$, as a first step to address this issue we re-analyzed the data in Fig.~\ref{fig:excdep}, by extracting only the single spectra $S(\nu,t_0)$ from the full spectrogram.  The resultant fitted values of $\taus$ are also included in Fig.~\ref{fig:excdep}(b) (solid triangles), and are indeed systematically larger than those obtained with the full spectrogram marginals (and then are  consistent with the value in Fig.~\ref{fig:tausgram}(c) for $\Nex=2.1\cdot 10^{19}\pcmc$).  Hence an important finding is that SF time-gating introduces artifacts in such measurements; we explain the reason for this below with the help of simulation results.

We modeled the experiments using 1D FDTD simulations, with a Drude response and spatio-temporal profile
for the photoinduced plasma (as described previously \cite{Thomson13,Meng14a}, except here with co-propagating pump-probe pulses and front-face excitation).
Simulation batches were carried out vs. the pump-probe delay $\tau$, using a time-independent nominal scattering time of $\tausz=25\fsec$. The numerical reflected fields were then analyzed to produce either the full spectrum or that corresponding to SF detection at a fixed time $t_0$ (with a 150-fs SF detection gate), corresponding to the various experimental results here.  
These spectra were then fitted using the same Drude model algorithm.
We present the results for three sets of simulation conditions in order to isolate various aspects expected in the analysis of the experimental data, as shown in Fig.~\ref{fig:tausgram}(d) (curves (i-iii)).

In the first case (i), we used a transform-limited (TL) THz-MIR probe pulse with an intensity spectrum based on the experimental one (Fig.~\ref{fig:refsgram}), a pump pulse duration equal to the experimentally observed $\usr{T}{r}$ for $\Nex$ in Fig.~\ref{fig:tausgram}(c), and fitted the full spectra of the numerical reflected THz-MIR pulses for each $\tau$.
The kinetics of the fitted $\Nex$-values are as expected and correspond closely to the experimental data.
Around zero-delay, the fitted values of $\taus$ are also seen to rise from $\lesssim 15\fsec$ toward the simulation value of $\tausz=25\fsec$ (note the slight error is due to the finite accuracy of the FDTD simulations).
However, it is also apparent that this artificial rise follows the same delay-dependence as the $\Nex$-curve, i.e. the effect occurs only during the formation of the plasma when there is temporal overlap between pump and probe pulses.

The second case (ii) adds the additional effects of our measurements in Fig.~\ref{fig:tausgram}(a-c), by adding the experimental chirp ($\varphi''=126\fsec^2$) to the input THz-MIR pulse, and using the numerically time-gated SF spectra.  This leads again to a rise in the fitted $\taus$-curves, albeit with a more complex variation during the leading portion of the excitation pulse, $\tau\lesssim 0$. (Note the fitted $\Nex$-curve is essentially the same for all simulation cases, and only that for case (i) is shown). However, in this case, one sees that fitted $\taus$ values in the steady-state are systematically larger than the underlying value of $\tausz$ used in the simulations (as seen in the experimental analysis above).  
This is not the case when the full spectra are used (with either TL or chirped  THz-MIR pulses, latter results not shown).  This allows us to identify the origin of the discrepancy above: the effect of the finite SF-detection gate pulse introduces a small distortion on the measured spectra which results in a systematic error (overestimation) in the fitted value of $\taus$, despite that fact that the gate is also applied to the detection of reference pulse (without plasma excitation).   
Having the detailed simulation results at hand, this can be traced to the small group-delay variation experienced by the THz-MIR pulse in reflection (see inset in Fig.~\ref{fig:excdep}(a)). While this is only of the order of $\pm 10\fsec$, it is indeed sufficient to distort the spectra in this way.
This is an important result, as one may have expected that the use of a reflection geometry should remove any influence of plasma dispersion (compared to bulk delays in transmission), and emphasizes the role of the finite group-delay in reflection from (and associated penetration into) a Drude medium.  
In future investigations this effect could be alleviated by the use of a significantly longer SF detection pulse (i.e. $\usr{T}{SF}\rightarrow 1\psec$ via spectral filtering).
Nevertheless, one sees that rise of the fitted $\taus$-curve for case (ii) still is essentially complete after the pump pulse excitation, and that these effects alone do not account for the slower rise in the experimental data (Fig.~\ref{fig:tausgram}(c)).

Now that we have demonstrated numerically that such experiments inherently produce artifacts in the fitted value of $\taus$ around zero-delay (which could cause physical misinterpretation of data, even with superior time-resolution), the question of its specific cause remains. About zero-delay there are two deviations from the steady-state situation assumed in the Drude model here (besides the possibility of a time-dependent $\taus$-value), i.e. not only does the THz-MIR pulse co-propagate with the pump pulse and experience a time-nonstationary medium, but also the excitation profile assumed in the model (exponential decay with $z$) is not yet fully formed and hence the expression used for calculating $r(\nu)$ (see Sec.~\ref{sec:excdep}, \cite{Vine1984}) is not completely accurate.
In order to isolate these two factors, we also performed a simulation batch (Fig.~\ref{fig:tausgram}(d), case (iii)) with a control situation where the plasma for each value of delay is kept artificially ``frozen'' at its profile for $t=\tau$ (and the THz-MIR pulse reflects from a time-\textit{stationary}, partially formed plasma profile).
As can be seen from the fitted $\taus$-curve (iii), while one still observes an artifact rise (due to the inaccuracy of the plasma profile model), this is now significantly more abrupt.  Hence one may conclude that the co-propagation and time-nonstationary medium dominate the slower rise seen in the curve for case (i).  This is an important result for interpretation of future experiments, as this effect will be present even for much shorter pump and probe pulses, and one should also employ detailed simulations of the co-propagation to avoid misinterpretation of experimental data.

Finally, returning to the delay-dependence of $\taus$ in Fig.~\ref{fig:tausgram}(c), in Fig.~\ref{fig:tausgram}(c) we provide a fit of the data (for $\tau\geq 200\fsec$) with an exponential step function $\propto (1-e^{-\tau/\usr{T}{s}})$ which yields a value of $\usr{T}{s}=450\fsec$. 
Given the simulation results above, it is clearly difficult to define and deconvolve a response function from the $\taus$-curve, and attempts to fit the $\taus$-curve including convolution with a Gaussian response degraded the quality of the fit.
In comparing this value of $\usr{T}{s}$ with the fitted Gaussian response for $\Nex$, one should consider the half-width-half-maximum of $\usr{T}{r}/2=210\fsec$.
Still, this uncorrected value of $\usr{T}{s}$ is close to $\usr{T}{s}=500\fsec$ reported in Ref.~\onlinecite{Ziel1996} (i.e. for a lower density, $\Nex=3 \cdot 10^{15}\pcmc$), and quite likely corresponds to subsequent energy relaxation of the photoexcited carriers.

\section{Conclusions}
The use of THz-MIR spectroscopy with a detailed quantitative analysis has yielded a more comprehensive estimate of the Drude scattering time of charge-carrier plasmas in undoped Si and its dependence on $\Nex$. 
The momentum relaxation time of the thermalized plasma ranges from $\simx 200\fsec$ at low electron-hole densities to $\simx 20\fsec$ for $\Nex\simx 10^{19}\pcmc$, at room temperature.
The direct comparison with theoretical predictions provides mutual support for both, and supports the mechanism of scattering rate saturation due to phase-space restrictions at high density. A review of the literature reports indicates that there are still open issues to address, both experimentally and theoretically. In particular, (i) the effect of excitation energy and temperature (in order to account for the differences between our results and those from e.g. Ref.~\onlinecite{Hend2007}), (ii) the magnitude of $\taus$ at very high densities (where either a revision of the experimental results, or additional physical scattering mechanisms, would be needed to reconcile the reported values $\taus\simx 1\fsec$), and (iii) the precise sub-picosecond plasma dynamics.  In the latter case, state-of-the-art theoretical methods to describe the nonequilibrium dynamics \cite{Bind95} are required. Our ongoing THz-MIR studies with improved time resolution are aimed at providing an experimental test for such theoretical predictions.
The results from auxiliary 1D-FDTD simulations revealed important aspects for conducting a reliable analysis of time-resolved spectra, and demonstrate that the spatio-temporal propagation effects must be taken into account even in a reflection geometry.

% If you have acknowledgments, this puts in the proper section head.
%\begin{acknowledgments}
%\end{acknowledgments}

% Create the reference section using BibTeX:
%\newpage
\bibliography{silicon_bib}

\end{document}